\documentclass[12pt]{article}%
\usepackage{amsmath,latexsym}
\usepackage{graphicx}
\usepackage{amsmath}
\usepackage{amsfonts}
\usepackage{amssymb}%
\begin{document}

\title{{The effect of conformal symmetry on charged
wormholes}}
   \author{
Peter K F Kuhfittig*\\
\footnote{E-mail: kuhfitti@msoe.edu}
 \small Department of Mathematics, Milwaukee School of
Engineering,\\
\small Milwaukee, Wisconsin 53202-3109, USA}

\date{}
 \maketitle

\begin{abstract}\noindent
This paper discusses the effect that conformal
symmetry can have on a charged wormhole.  The
analysis yields a physical interpretation of
the conformal factor in terms of the electric
charge.  The rate of change of the conformal
factor determines much of the outcome, which
ranges from having no solution to wormholes
having either one or two throats.   \\

\noindent
Keywords\\
Charged Wormholes, Conformal Symmetry\\
\end{abstract}

\section{Introduction}\label{S:Introduction}

Wormholes are handles or tunnels in spacetime
connecting different regions of our Universe
or different universes altogether.  That
wormholes could be actual physical structures
suitable for interstellar travel was first
proposed by Morris and Thorne \cite{MT88}.
Such wormholes can be described by the static
and spherically symmetric line element
\begin{equation}\label{E:line1}
ds^{2}=-e^{2\Phi(r)}dt^{2}+\frac{dr^2}{1-b(r)/r}
+r^{2}(d\theta^{2}+\text{sin}^{2}\theta\,
d\phi^{2}),
\end{equation}
using units in which $c=G=1$.  Here
$\Phi=\Phi(r)$ is called the \emph{redshift
function}, which must be everywhere finite
to avoid an event horizon.  The function
$b=b(r)$ is called the \emph{shape function}
since it determines the spatial shape of the
wormhole when viewed, for example, in an
embedding diagram \cite{MT88}.  The spherical
surface $r=r_0$ is the \emph{throat} of the
wormhole.  Here $b(r)$ must satisfy the
following conditions: $b(r_0)=r_0$, $b(r)<r$
for $r>r_0$, and  $b'(r_0)\le 1$, usually called
the \emph{flare-out condition}.  For
Morris-Thorne wormholes, this condition
can only be satisfied by violating the null
energy condition, requiring the use of
``exotic matter."  Since $b'(r)$ is
proportional to the density in the Einstein
field equations, we ordinarily require that
$b'(r)>0$.

In this paper we study the effect of conformal
symmetry on wormholes that have an electric
charge.  More precisely, we assume the existence
of a conformal Killing vector $\xi$ defined by
the action of $\mathcal{L_{\xi}}$ on the metric
tensor:
\begin{equation}
  \mathcal{L_{\xi}}g_{\mu\nu}=\psi(r)\,g_{\mu\nu},
\end{equation}
where $\mathcal{L_{\xi}}$ is the Lie derivative
operator and $\psi(r)$ is the conformal factor.
Charged wormholes were first proposed by Kim
and Lee \cite{KL01}.  Compatibility of such
wormholes with quantum field theory is discussed
in Ref. \cite{pK11}.

In addition to studying its effect on a charged
wormhole, we obtain a physical interpretation of
the conformal factor in terms of the electric
charge.  The combination of electric charge and
conformal  symmetry results in a wormhole model
that may actually have two throats.

\section{Conformal Killing vectors}\label{S:Killing}
    \label{S:Killing}
As noted in the Introduction, we assume that our
static spherically symmetric spacetime admits a
one-parameter group of conformal motions, i.e.,
motions along which the metric tensor remains
invariant up to a scale factor.  Equivalently,
there exist conformal  Killing vectors such
that
\begin{equation}\label{E:Lie}
   \mathcal{L_{\xi}}g_{\mu\nu}=g_{\eta\nu}\,\xi^{\eta}
   _{\phantom{A};\mu}+g_{\mu\eta}\,\xi^{\eta}_{\phantom{A};
   \nu}=\psi(r)\,g_{\mu\nu},
\end{equation}
where the left-hand side is the Lie derivative of the
metric tensor and $\psi(r)$ is the conformal factor
\cite{MM96, BHL07}.  The vector $\xi$ generates the
conformal symmetry and the metric tensor $g_{\mu\nu}$ is
conformally mapped into itself along $\xi$.  As
discussed in Refs. \cite{HPa, HPb}, this type of symmetry
has been used effectively to describe relativistic
stellar-type objects, thereby leading to new solutions,
as well as new geometric and kinematical insights
\cite{MS93, Ray08, fR10, fR12b, BHL08}.  Even more
significantly, it has also been shown that the Kerr
black hole is conformally symmetric \cite{CMS10}.
Two earlier studies assumed \emph{non-static}
conformal symmetry \cite{BHL07, BHL08}.

To study the effect of conformal symmetry, it is
convenient to use an alternate form of the metric
\cite{RRKKI, pK16}:
\begin{equation}\label{E:line2}
   ds^2=- e^{\nu(r)} dt^2+e^{\lambda(r)} dr^2
   +r^2( d\theta^2+\text{sin}^2\theta\, d\phi^2).
\end{equation}
Using this form, the Einstein field equations
become
\begin{equation}\label{E:Einstein1}
e^{-\lambda}
\left(\frac{\lambda^\prime}{r} - \frac{1}{r^2}
\right)+\frac{1}{r^2}= 8\pi \rho,
\end{equation}

\begin{equation}\label{E:Einstein2}
e^{-\lambda}
\left(\frac{1}{r^2}+\frac{\nu^\prime}{r}\right)-\frac{1}{r^2}=
8\pi p_r,
\end{equation}

\noindent and

\begin{equation}\label{E:Einstein3}
\frac{1}{2} e^{-\lambda} \left[\frac{1}{2}(\nu^\prime)^2+
\nu^{\prime\prime} -\frac{1}{2}\lambda^\prime\nu^\prime +
\frac{1}{r}({\nu^\prime- \lambda^\prime})\right] =8\pi p_t.
\end{equation}

To keep the analysis tractable, we follow Herrera and
Ponce de Le\'{o}n \cite{HPa} and restrict the vector field
by requiring that $\xi^{\alpha}U_{\alpha}=0$, where
$U_{\alpha}$ is the four-velocity of the perfect fluid
distribution, so that fluid flow lines are mapped
conformally onto fluid flow lines.  The assumption
of spherical symmetry then implies that
$\xi^0=\xi^2=\xi^3=0$ \cite{HPa}.  Eq. (\ref
{E:Lie}) now yields the following results:
\begin{equation}\label{E:sol1}
    \xi^1 \nu^\prime =\psi,
\end{equation}
\begin{equation}\label{E:sol2}
   \xi^1  = \frac{\psi r}{2},
\end{equation}
and
\begin{equation}\label{E:sol3}
  \xi^1 \lambda ^\prime+2\,\xi^1 _{\phantom{1},1}=\psi.
\end{equation}
From Eqs. (\ref{E:sol1}) and (\ref{E:sol2}),
we then obtain $\nu'=2/r$ and thereby
\begin{equation} \label{E:gtt}
   e^\nu  =c_1 r^2,
\end{equation}
where $c_1$ is an integration constant.  Now from
Eq. (\ref{E:sol2}) we get
\[
   \xi^1 _{\phantom{1},1}=\frac{1}{2}
   (\psi'r+\psi).
\]
Substituting in Eq. (\ref{E:sol3}) and using
$\nu'=2/r$, simplification yields
\[
     \lambda'=-2\frac{\psi'}{\psi}.
\]
Solving for $\lambda$ produces the final result,
\begin{equation}\label{E:grr}
   e^\lambda  = \left(\frac {c_2} {\psi}\right)^2,
\end{equation}
where $c_2$ is another integration constant.

The Einstein field equations can be
rewritten as follows:
\begin{equation}\label{E:E1}
\frac{1}{r^2}\left(1 - \frac{\psi^2}{c_2^2}
\right)-\frac{(\psi^2)^{\prime}}{c_2^2r}= 8\pi \rho,
\end{equation}
\begin{equation}\label{E:E2}
\frac{1}{r^2}\left( \frac{3\psi^2}{c_2^2}-1
\right)= 8\pi p_r,
\end{equation}
and
\begin{equation}\label{E:E3}
\frac{\psi^2}{c_2^2r^2}
+\frac{(\psi^2)^{\prime}}{c_2^2r} =8\pi p_t.
\end{equation}
It now becomes apparent that $c_2$ is merely a
scale factor in Eqs. (\ref{E:grr})-(\ref{E:E3});
so we may assume that $c_2=1$.  The constant
$c_1$, on the other hand, will have to be
obtained from the junction conditions, the
need for which can be seen from Eq.
(\ref{E:gtt}):
since our wormhole spacetime is not asymptotically
flat, the wormhole material must be cut off at
some $r=a$ and joined to an exterior
Schwarzschild solution,
\begin{equation}\label{E:line3}
ds^{2}=-\left(1-\frac{2M}{r}\right)dt^{2}
+\frac{dr^2}{1-2M/r}
+r^{2}(d\theta^{2}+\text{sin}^{2}\theta\,
d\phi^{2}),
\end{equation}
so that $e^{\nu(a)}=c_1a^2=1-2M/a$, whence
\begin{equation}\label{E:junction}
    c_1=\frac{1-2M/a}{a^2},
\end{equation}
where $M$ is the mass of the wormhole as
seen by a distant observer.

\section{Charged wormholes}

It was proposed by Kim and Lee \cite{KL01} that
for a wormhole with constant charge $Q$ the
Einstein field equations take on the form
\begin{equation}\label{E:EFE}
   G^{(0)}_{\mu\nu}+G^{(1)}_{\mu\nu}=8\pi
      [T^{(0)}_{\mu\nu}+T^{(1)}_{\mu\nu}].
\end{equation}
Given that the usual form is $G^{(0)}_{\mu\nu}=
8\pi T ^{(0)}_{\mu\nu}$, the modified form in
Eq. (\ref{E:EFE}) is obtained by adding the matter
term $T^{(1)}_{\mu\nu}$ to the right side  and
the corresponding back reaction term
$G^{(1)}_{\mu\nu}$ to the left side.  The proposed
metric is
\begin{equation}\label{E:line4}
  ds^2=-\left(1+\frac{Q^2}{r^2}\right)dt^2
   +\left(1-\frac{b(r)}{r}+\frac{Q^2}{r^2}\right)^{-1}dr^2\\
    +r^2(d\theta^2+\text{sin}^2\theta\,d\phi^2).
\end{equation}
Kim and Lee go on to note that with $b\equiv 0$, the
wormhole becomes a Reissner-Nordstr\"{o}m black hole,
and if $Q=0$, the spacetime becomes a Morris-Thorne
wormhole with a shape function $b=b(r)$ that meets
the usual requirements.  It therefore became necessary
to show that the metric, Eq. (\ref{E:line4}), is a
self-consistent solution of the Einstein field
equations.  The shape function $b=b(r)$ of the
Morris-Thorne wormhole is now replaced by the
effective shape function
\begin{equation}\label{E:beff1}
    b_{\text{eff}}(r)=b(r)-\frac{Q^2}{r}.
\end{equation}
The effective shape function also has the usual
properties, to be discussed later.

\section{Charged wormholes with conformal symmetry}

In this section we return to the assumption of
conformal symmetry mentioned in Sec. \ref{S:Killing}.
Let us first consider the Kim-Lee model, Eq.
(\ref{E:line4}).  Then by Eq. (\ref{E:gtt}),
$e^{\nu}=c_1r^2$, we have, for all $r$,
\[
     1+\frac{Q^2}{r^2}=c_1r^2,
\]
which is impossible.  So this model is not
compatible with the assumption of conformal
symmetry.  This difficulty can be overcome,
however, by introducing a new differentiable
function $S(r)$ to yield the line element
\begin{equation}\label{E:line5}
  ds^2=-\left(1+S(r)+\frac{Q^2}{r^2}\right)dt^2\\
   +\left(1-\frac{b(r)}{r}
   +\frac{Q^2}{r^2}\right)^{-1}dr^2
    +r^2(d\theta^2+\text{sin}^2\theta\,d\phi^2).
\end{equation}
Evidently,
\begin{equation}
   S(r)=-\left(1+\frac{Q^2}{r^2}\right)+
   r^2\frac{1-2M/a}{a^2}
\end{equation}
by Eq. (\ref{E:junction}).  Since $a>r$ on the
interval $[r_0,a]$, it follows that $S(r)<0$,
while $S'(r)>0$.

As already noted, given the effective shape
function $b_{\text{eff}}(r)=b(r)-Q^2/r$
and the total matter $T^{\text{eff}}=
T_{\mu\nu}^{(0)}+T_{\mu\nu}^{(1)},$ the
Kim-Lee model yields a self-consistent
solution.  The inclusion of $S(r)$ has no
effect on this conclusion.  So our metric,
Eq. (\ref{E:line5}), is a valid solution of
the Einstein field equations representing a
wormhole with an electric charge.

The major objective in this section is to
obtain a physical interpretation of the
conformal factor $\psi (r)$, as well as the
restrictions required to obtain a wormhole.
First we recall that for some $r=r_1$,
$b(r_1)=r_1$ and $b'(r_1)\le 1$.  Also, for
$r>r_1$, $b(r)<r$.  For the new (effective)
shape function, $b_{\text{eff}}(r)=b(r)-Q^2/r$,
we have $b_{\text{eff}}(r_0)=r_0$ and
$b'_{\text{eff}}(r_0)\le 1$ to meet the
flare-out condition.  (This implies that
$b'(r_0)\le 1-Q^2/r_0^2$.)  Since $b=b(r)$
is assumed to be a typical shape function,
$b(r)<r$ for $r>r_1$, and $b(r)>r$ for
$r<r_1$.  Since $b(r_0)-r_0=Q^2/r_0>0$ by
Eq. (\ref{E:beff1}), it follows that
$r_0<r_1$.  (See Fig. 1.)
\begin{figure}[tbp]
\begin{center}
\includegraphics[width=0.8\textwidth]{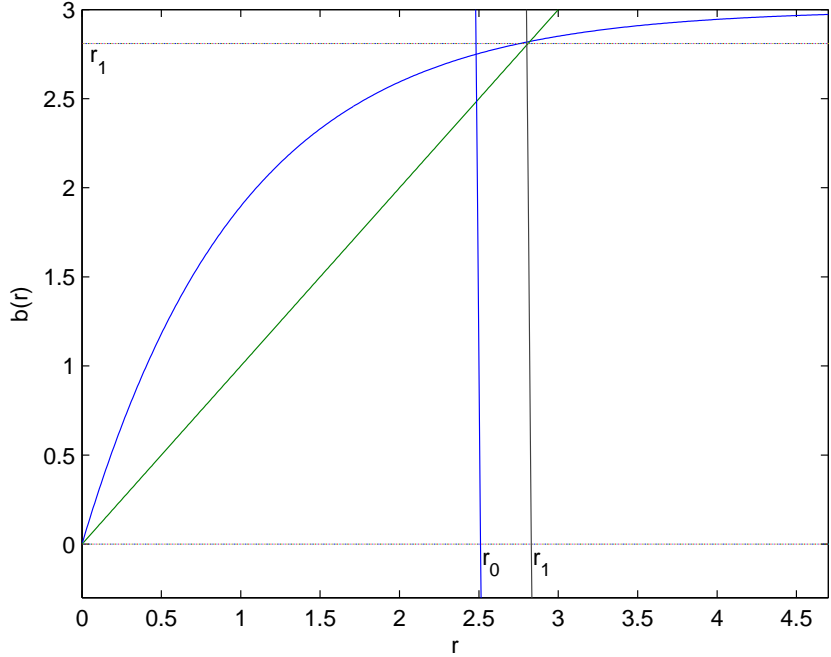}
\end{center}
\caption{The throat of $b_{\text{eff}}(r)$ at
    $r=r_0$ remains near $r=r_1$.}
\end{figure}

Next, from Eqs. (\ref{E:line2}), (\ref{E:grr}),
and (\ref{E:line5}) (and recalling that $c_2=1$),
\begin{equation}\label{E:psishape}
   1-\frac{b(r)}{r}+\frac{Q^2}{r^2}=\psi^2(r).
\end{equation}
Since $b(r)/r=1+Q^2/r^2-\psi^2(r)$, it also
follows that
\begin{equation}\label{E:shape1}
   b(r)=r\left(1+\frac{Q^2}{r^2}-
   \psi^2(r)\right)
\end{equation}
and
\begin{equation}
   b_{\text{eff}}(r)=r
   \left(1+\frac{Q^2}{r^2}-\psi^2(r)-
   \frac{Q^2}{r^2}\right)=r(1-\psi^2(r)).
\end{equation}
The condition $b_{\text{eff}}(r_0)=r_0$ now
implies that
\begin{equation}\label{E:psi1}
   \psi^2(r_0)=0.
\end{equation}
Also, since $b(r_1)=r_1$, we have
\[
   1-\frac{b(r_1)}{r_1}+\frac{Q^2}{r_1^2}
   =\psi^2(r_1)
\]
and
\begin{equation}\label{E:psi2}
   \psi^2(r_1)=\frac{Q^2}{r_1^2}.
\end{equation}
So Eqs. (\ref{E:psi1}) and (\ref{E:psi2}) give
us a physical interpretation for the conformal
factor $\psi(r)$ in terms of the charge $Q$.
Additional connections to $Q$ are given below.

Having just learned that $\psi^2(r)$ increases
to the right of $r=r_0$, let us examine the
slope of $\psi^2(r)$ more closely.  First note
that from Eq. (\ref{E:shape1}) we have
 \begin{equation}\label{E:bprime}
    \left.b'(r)=1-\psi^2(r)-r\frac{d}{dr}\psi^2(r)
    \right|_{r=r_0}=1-r_0\frac{d}{dr}\psi^2(r_0).
 \end{equation}
From the inequality $0<b'_{\text{eff}}(r_0)\le 1$,
we get
\[
   0<b'(r_0)+\frac{Q^2}{r_0^2}=1-r_0\frac{d}{dr}
   \psi^2(r_0)+\frac{Q^2}{r_0^2}\le 1.
\]
Solving, we obtain the inequality
\begin{equation}\label{E:ineq1}
   \frac{Q^2}{r_0^3}\le\frac{d}{dr}\psi^2(r_0)
   <\frac{1}{r_0}+\frac{Q^2}{r_0^3}.
\end{equation}

Similarly, since $\psi^2(r_1)=Q^2/r_1^2$ and
$0<b'(r_1)\le 1$, we have
\[
   0<1-\frac{Q^2}{r_1^2}-r_1\frac{d}{dr}
   \psi^2(r_1)\le 1.
\]
Solving, we obtain the second inequality:
\begin{equation}\label{E:ineq2}
   -\frac{Q^2}{r_1^3}\le\frac{d}{dr}
   \psi^2(r_1)<\frac{1}{r_1}-\frac{Q^2}{r_1^3}.
\end{equation}

Our final task is to check the violation of the
null energy condition (NEC) required to hold the
wormhole open.  Recall that the NEC states that
given the stress-energy tensor $T_{\alpha\beta}$,
\[
   T_{\alpha\beta}\mu^{\alpha}\mu^{\beta}\ge 0
\]
for all null vectors.  So we obtain from the null
vector $(1,1,0,0)$ that $\rho +p_r<0$ whenever the
condition is violated.  By Ref. \cite{MT88}, this
violation is equivalent to the condition
\[
   \frac{b'(r_0)-b(r_0)/r_0}{2[b(r_0)]^2}<0
\]
for a generic shape function.  For
$b_{\text{eff}}(r)$, we already know that the
last inequality holds whenever $\frac{d}{dr}
\psi^2(r_0)\ge Q^2/r_0^3$ by inequality
(\ref{E:ineq1}).  By Eq. (\ref{E:psishape}),
$\psi^2(r)$ is associated with $b_{\text{eff}}$;
so we can use the field equations, Eqs.
(\ref{E:E1}) and (\ref{E:E2}), to determine
\begin{equation}
   \left.8\pi (\rho +p_r)\right|_{r=r_0}=
   -\frac{1}{r_0}\frac{d}{dr}\psi^2(r_0)
\end{equation}
since $\psi^2(r_0)=0$.  Near the throat $r=r_0$,
$\psi^2(r)$ is rising, so that the NEC is indeed
violated at and near $r=r_0$.

For $r=r_1$, the throat of $b=b(r)$, we cannot
use Eqs. (\ref{E:E1}) and (\ref{E:E2}), since
the null vectors are not the same.  Moreover,
we can infer from Eq. (\ref{E:bprime}) that
\begin{equation}
   0<\psi^2(r)+r\frac{d}{dr}\psi^2(r)<1,
\end{equation}
showing that $\psi^2(r)$ cannot keep increasing
indefinitely.  In fact, given that $\frac{d}{dr}
\psi^2(r_1)\ge -Q^2/r_1^3$, $\psi^2(r)$ could
already be decreasing at $r=r_1$.  We therefore
have to require that $r=r_1$ be close enough
to $r=r_0$ so that
\begin{equation}
   \frac{d}{dr}\psi^2(r_1)\ge 0.
\end{equation}
(Since $Q^2$ is small in geometrized units, $r_1$
is close to $r_0$ to begin with.)  As a result,
if the NEC is violated at $r=r_0$, it is also
violated at $r=r_1$.

\section{An analogue of the Kerr-Newman black hole}

In thia section we study the various conditions
under which the NEC is violated.  So let us restate
Inequality (\ref{E:ineq1}) and the modified
Inequality (\ref{E:ineq2}):
\begin{equation}\label{E:first}
   \frac{Q^2}{r_0^3}\le\frac{d}{dr}\psi^2(r_0)<
   \frac{1}{r_0}+\frac{Q^2}{r_0^3}
\end{equation}
and
\begin{equation}\label{E:second}
   0\le\frac{d}{dr}\psi^2(r_1)<\frac{1}{r_1}
   -\frac{Q^2}{r_1^3}.
\end{equation}
From
\[
   0<\frac{Q^2}{r_0^3}<\frac{1}{r_1}-
   \frac{Q^2}{r_1^3}<\frac{1}{r_0}
   +\frac{Q^2}{r_0^3},
\]
we obtain the following half-open intervals:
\[
   \text{Interval I}:\quad
   \left.\left[0,\frac{Q^2}{r_0^3}
   \right)\right.,
\]
\[
   \text{Interval II}:\quad
   \left.\left[\frac{Q^2}{r_0^3},
   \frac{1}{r_1}-\frac{Q^2}{r_1^3}
   \right)\right.,
\]
and
\[
   \text{Interval III}:\quad
   \left.\left[\frac{1}{r_1}-\frac{Q^2}{r_1^3},
   \frac{1}{r_0}+\frac{Q^2}{r_0^3}
   \right)\right..
\]

We now observe that whenever $\frac{d}{dr}
\psi^2(r)$ is in Interval I, the NEC is
violated at $r=r_1$, but not at $r=r_0$.
(The reason is that $\frac{d}{dr}
\psi^2(r_1)$ now satisfies Inequality
(\ref{E:second}), but $\frac{d}{dr}
\psi^2(r_0)$ does not satisfy Inequality
(\ref{E:first}).)  So only $b(r)$ has a
legitimate throat.  If
$\frac{d}{dr}\psi^2(r)$ is in Interval III,
the NEC is violated at $r=r_0$, but not at
$r=r_1$.  (Here $\frac{d}{dr}
\psi^2(r_0)$ satisfies Inequality
(\ref{E:first}), but $\frac{d}{dr}
\psi^2(r_1)$ does not satisfy Inequality
(\ref{E:second}).)  So only
$b_{\text{eff}}(r)$ has a legitimate throat.
Finally, if $\frac{d}{dr} \psi^2(r)$ is in
Interval II, the NEC is violated at both
$r_0$ and $r_1$.  So given the right
conditions, our wormhole can have two
throats.

Now recall that the event horizon of a black
hole is often viewed as the analogue of the
throat of the wormhole.  In fact, according
to Hayward \cite{sH02}, if enough negative
energy is injected into a black hole, it may
become a traversable wormhole; the event
horizon becomes the throat.  From this
perspective, our wormhole can be viewed
as the natural analogue of the Kerr-Newman
black hole: this type of black hole also
has two surfaces that are characterized
by coordinate singularities.

\emph{Remark:} The existence of two throats
invites the following speculation: a variation
on the Kim-Lee model is
\begin{equation}
  ds^2=-e^{\Phi(r,Q)}dt^2
   +\left(1-\frac{b(r)}{r}
   +\frac{Q^2}{r^2}\right)^{-1}dr^2
    +r^2(d\theta^2+\text{sin}^2\theta\,d\phi^2).
\end{equation}
Unlike our earlier model, this metric can lead to
an event horizon.  In particular, suppose
\[
   e^{\Phi(r,Q)}=e^{-Q^2/(r-r_2)^2},
\]
where $r_0<r_2<r_1$.  Since Inequalities
(\ref{E:first}) and (\ref{E:second}) still hold,
this model suggests that it is in principle
possible to pass through the throat at
$r=r_1$ and return via the throat at $r=r_0$,
not only bypassing the event horizon of the
black hole, but allowing a return trip.

\section{Conclusion}

This paper discusses the effect that conformal
symmetry can have on a charged wormhole.
Conversely, the physical requirements are seen
to place severe constraints on the wormhole
geometry.

The analysis yields a physical interpretation of
the conformal factor $\psi(r)$ in terms of the
charge $Q$.  Moreover, the outcome is heavily
dependent on $\psi^2(r)$ and
$\frac{d}{dr}\psi^2(r)$ and ranges from
having no solution to wormholes having two
throats.  The latter case can be viewed as
the analogue of the Kerr-Newman black hole.

\end{document}